\documentclass[10pt,letterpaper]{article}
\usepackage{opex3}
\usepackage{subfig}
\usepackage{amsmath}
\usepackage{cite}

\begin{document}

\title{Design of thin--film photonic metamaterial L\"uneburg lens using analytical approach}

\author{Hanhong Gao,$^{1,*}$ Baile Zhang,$^{2,3}$ Steven G. Johnson,$^4$ and George Barbastathis$^{2,5}$}

\address{
$^1$Department of Electrical Engineering and Computer Science, Massachusetts Institute of Technology, \\ 77 Massachusetts Avenue, Cambridge, MA 02139, USA\\
$^2$Singapore-MIT Alliance for Research and Technology (SMART) Centre, \\ 3 Science Drive 2, Singapore 117543, Singapore \\
$^3$Division of Physics and Applied Physics, School of Physical and Mathematical Sciences, \\ Nanyang Technological University, Singapore 637371, Singapore\\
$^4$Department of Mathematics, Massachusetts Institute of Technology, \\ 77 Massachusetts Avenue, Cambridge, MA 02139, USA \\
$^5$Department of Mechanical Engineering, Massachusetts Institute of Technology, \\ 77 Massachusetts Avenue, Cambridge, MA 02139, USA
}

\email{gaohh87@mit.edu} 



\begin{abstract}
We design an all--dielectric L\"uneburg lens as an adiabatic space--variant lattice explicitly accounting for finite film thickness. We describe an all-analytical approach to compensate for the finite height of subwavelength dielectric structures in the pass--band regime. This method calculates the effective refractive index of the infinite--height lattice from effective medium theory, then embeds a medium of the same effective index into a slab waveguide of finite height and uses the waveguide dispersion diagram to calculate a new effective index. The results are compared with the conventional numerical treatment -- a direct band diagram calculation, using a modified three--dimensional lattice with the superstrate and substrate included in the cell geometry. We show that the analytical results are in good agreement with the numerical ones, and the performance of the thin--film L\"uneburg lens is quite different than the estimates obtained assuming infinite height.
\end{abstract}



\section{Introduction}

Gradient Index (GRIN) media have been known to offer rich possibilities for light manipulation since at least Maxwell's time \cite{Maxwell1854}. More recent significant examples are the L\"uneburg lens \cite{Luneburg1944}, the Eaton lens \cite{Eaton1953}, and the plethora of imaging and cloaking configurations devised recently using conformal maps and transformation optics \cite{Leonhardt2006,Pendry2006,Valentine2009,Vasic2010,Spadoti2010}. GRIN optics are of course also commercially available, but the achievable refractive index profiles $n(\mathbf{r})$ are limited generally to parabolic in the lateral coordinates or to axial without any lateral dependence \cite{Hecht2002}. There is an ongoing effort to achieve more general distributions using stacking of photo-exposed polymers \cite{Schmidt2009,DARPA2010}.

For optics--on--a--chip or integrated optics applications, it is possible to {\em emulate} an effective index distribution $n(\mathbf{r})$ by patterning a substrate with subwavelength structures. If these are sufficiently smaller than the wavelength, to a good approximation they can be thought of as a continuum where the effective index is determined by the pattern geometry. For example, one can create a lattice of alternating dielectric--air with slowly varying period and fixed duty cycle, or with fixed period but slowly varying duty cycle \cite{Jiao2004,Russel1999}.

If the critical length of the variation is slow enough compared to the lattice constant that the adiabatic approximation is valid, the lattice dispersion diagram can be used to estimate the local effective index \cite{Jiao2004,Russel1999}. Refractive indices computed using a 2D approximation are valid for 2D adiabatically variant metamaterials where the height in the $3^{\text{rd}}$ dimension is much larger than the wavelength so the assumption of infinite height can be justified. According to this, we have designed a subwavelength aperiodic nanostructured L\"uneburg lens \cite{Takahashi2010,Takahashi2011}. This lens mimics a GRIN element with refractive index distribution $n(\rho)=n_0\sqrt{2-(\rho/R)^2}~(0<\rho <R)$, where $n_0$ is the ambient index outside the lens region, $R$ is the radius of the lens region and $\rho$ is the radial polar coordinate with the lens region as origin. The L\"uneburg lens focuses an incoming plane wave from any arbitrary direction to a geometrically perfect focal point at the opposite edge of the lens \cite{Luneburg1944,Vakil2011}.

However, most such adiabatically variant structures are fabricated by etching holes or rods on a thin silicon film, whose height is less than even the optical wavelength \cite{Valentine2009,Gabrielli2009,Takahashi2010,Takahashi2011,Zentgraf2010}. Hence, the infinite height assumption becomes questionable. Moreover, the structures are asymmetric since typically beneath the structure there is a substrate such as glass, whereas above the structure is air. Asymmetry also induces a long-wavelength cutoff in the guided modes \cite{Lee2008}; therefore the thin-film metamaterial should operate in an intermediate regime where the wavelength is neither too large nor too small. The problem of asymmetry and finite height have been acknowledged in the literature on photonic crystals \cite{Ulrich1973,Meade1994,Fan1997,Johnson1999,Qiu2002}, where the most common solution is to compute a full 3D band diagram \cite{Joannopoulos2008}. Most of them focus on photonic crystal slabs operating at wavelengths comparable to the periodicity, discussing phenomena such as super-collimation \cite{Witzens2002}, negative refraction \cite{Ahmadlou2006}, etc. To the best of our knowledge, the same problem has received insufficient attention in the context of 2D dielectric periodic or aperiodic metamaterial devices, especially those operating at the propagation regime of the band diagram. It has been briefly mentioned in \cite{Valentine2009,Zhang2011} without giving a detailed solution.

In our fabricated L\"uneburg lens design, thin--film problem is obvious where the experimental results show dislocated and aberrated focal point \cite{Takahashi2010,Takahashi2011}. In this paper we re--designed the L\"uneburg lens to include the finite film thickness, improving the estimate of the expected focal point position. To design such a lens, first we need a method for estimating effective refractive index of thin--film metamaterials. Several methods have been proposed in the literature. A conventional numerical approach (we refer to it as Direct Band Diagram, DBD) in photonic crystals derives a 3D lattice cell from the original 2D cell by surrounding a finite--height rod with large spaces of air above and glass substrate below \cite{Joannopoulos2008}. Another method takes one unit cell and retrieve the refractive index by its reflection and refraction properties \cite{Chen2004}. These methods yield accurate results but require either 3D band or finite--difference calculations. More heuristic (but faster) effective--index methods estimate a slab--waveguide effective index first and then use it to compute a 2D band diagram or effective index \cite{Hammer2009}. They are generally suitable for structures with etched substrates. In contrast, our proposal essentially reverses the order of these steps: we compute an effective index from the 2D cross--section first, and then incorporate it into a slab--waveguide mode. This is more suitable to the metamaterial regime.

In particular, we propose the following all--analytical method for effective refractive index calculation. First, we replace the rods with a continuum of a certain effective permittivity $\epsilon_{\mathrm{eff}}^{\mathrm{2D}}$. We calculate $\epsilon_{\mathrm{eff}}^{\mathrm{2D}}$ from 2D lattice of infinite--height rods using second-order effective medium theory, and then substitute $\epsilon_{\mathrm{eff}}^{\mathrm{2D}}$ as the permittivity of a slab of finite thickness, acting as an effective guiding medium, sandwiched between semi-infinite spaces of air above and glass below. The geometry then becomes one of a weakly-guiding waveguide due to the small height of the effective guiding medium (See Fig.~\ref{fig:GuidanceCondition}). This weakly--guiding effect modifies the real part of the horizontal wave--vector component, and thus a new effective permittivity $\epsilon_{\mathrm{eff}}^{\mathrm{3D}}$ for the finite slab of rods is derived from the waveguide dispersion relationship. We refer to this method as Effective Guiding Medium (EGM). Comparing with rigorous 3D calculations, our method provides more physical insights, and is generally faster to compute.

To validate our method, we compare it with the DBD method. It is shown that the results of both methods are in good agreement.



\begin{figure}
\centering
\subfloat[]{ \includegraphics[width=4cm]{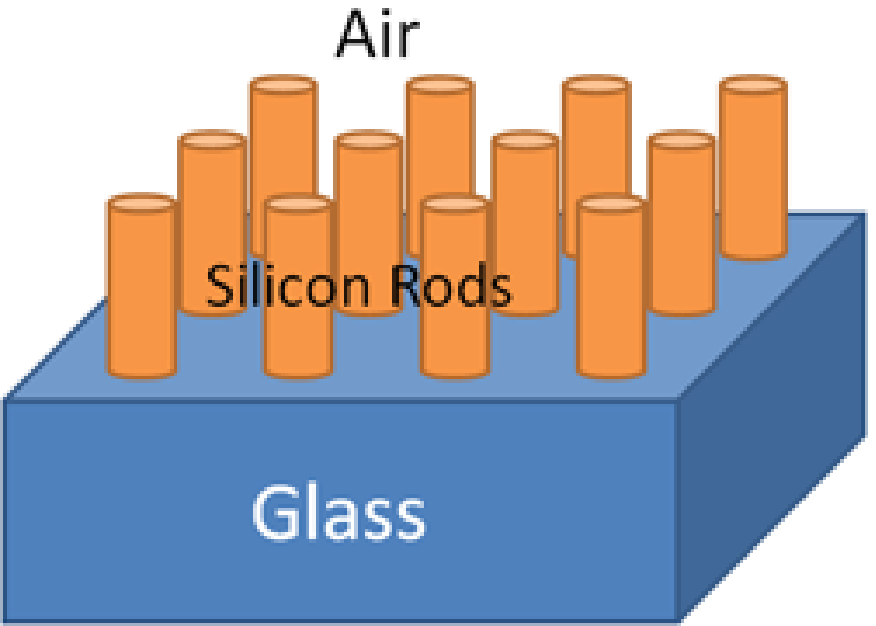} }
\subfloat[]{ \includegraphics[width=3.2cm]{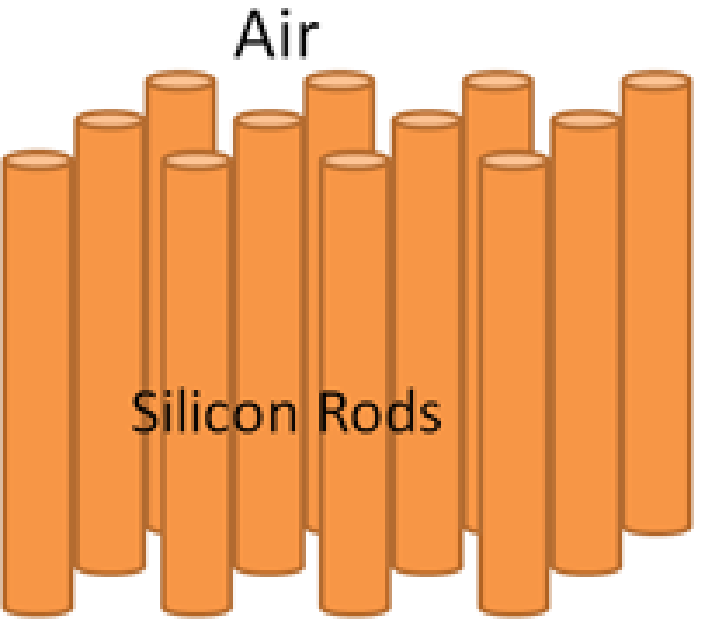} }
\caption{(a) Finite height rod lattice structure investigated in this paper. (b) 2D rod lattice structure assuming infinite height.}
\label{fig:struct}
\end{figure}

\section{Analytical method for effective refractive index estimation}

In this paper, without loss of generality, we investigate a silica glass slab covered by a square lattice (lattice constant $a=258~\mathrm{nm}$) of silicon rods of finite height $h=320~\mathrm{nm}$, variable radius $r$ ($0<r<a/\sqrt{2}$) and immersed in air, as illustrated in Fig.~\ref{fig:struct}(a). The free space wavelength of light is chosen as $\lambda=6a=1550~\mathrm{nm}$. This choice of $a$ is small enough to insure that we remain in the metamaterial regime and in the propagating regime of the band diagram; and large enough that the rods can be accurately fabricated by nano--lithography \cite{Takahashi2010,Takahashi2011} and we do not reach the long--wavelength cutoff regime for the asymmetric waveguide, as mentioned above. The dielectric permittivity constants for glass and silicon are $\epsilon_{\mathrm{glass}}=2.25$ and $\epsilon_{\mathrm{silicon}}=12.0$, respectively. These media are non--magnetic, so the relative permeability is taken as $\mu=1$ throughout this paper. The glass slab height is assumed to be much larger than the height of the rods and the free space wavelength of the light. The corresponding 2D structure with infinite height rods and without glass substrate is shown in Fig.~\ref{fig:struct}(b). We now proceed to describe all--analytical method, EGM, for analyzing these two geometries.

\subsection{Effective guiding medium (EGM) method}

\begin{figure}[htbp]
\centering
\includegraphics[width=10cm]{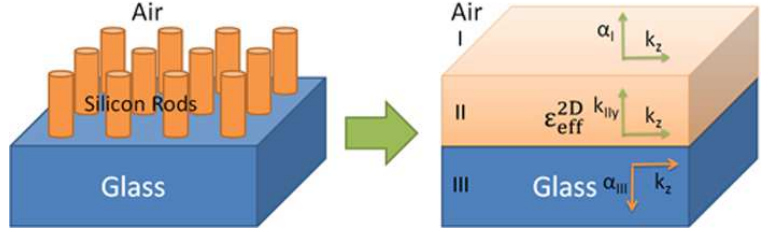}
\caption{Effective guiding medium (EGM) approximation of 2D finite height rod lattice structure.}
\label{fig:GuidanceCondition}
\end{figure}

The EGM method requires analysis of a three--layer structure: (I) air, (II) effective medium waveguide and (III) glass, as shown in Fig.~\ref{fig:GuidanceCondition}. The effective permittivity of the guiding medium is calculated from the second--order effective medium theory in 2D which have been derived by various authors \cite{Rytov1956,Brauer1994}. This theory starts from the effective refractive index of 1D subwavelength grating composed of air and dielectric with index $n$. Under TE (electric field parallel to the grooves) and TM (electric field vertical to the grooves) polarization incidence the effective index can be summarized, respectively, as \cite{Brauer1994,Yu2002}
\begin{eqnarray}
n_{\mathrm{TE}}^2 &=& n_{\mathrm{0TE}}^2+\frac{\pi^2}{3}\bigg(\frac{T}{\lambda}\bigg)f^2(1-f)^2(n^2-1)^2,
\label{eq:1DEffectiveIndex1}\\
n_{\mathrm{TM}}^2 &=& n_{\mathrm{0TM}}^2+\frac{\pi^2}{3}\bigg(\frac{T}{\lambda}\bigg)f^2(1-f)^2n_{\mathrm{0TM}}^6n_{\mathrm{0TE}}^2 \bigg(\frac{1}{n^2}-1\bigg)^2,
\label{eq:1DEffectiveIndex2}
\end{eqnarray}
where
\begin{eqnarray}
n_{\mathrm{0TE}}^2=fn^2+(1-f),~~~n_{\mathrm{0TM}}^2=1\bigg/\bigg(\frac{f}{n^2}+(1-f)\bigg)
\end{eqnarray}
are the zeroth-order effective refractive indices, $T$ is the period of the grating and $f$ is the filling factor of the dielectric grooves. The effective indices of corresponding 2D subwavelength structures are then estimated as a combination of 1D structures \cite{Brauer1994,Yu2002}
\begin{eqnarray}
n_{\mathrm{2D-TE}} &=& \sqrt{1-f+fn_{\mathrm{TE}}^2},\\
n_{\mathrm{2D-TM}} &=& \bigg( \sqrt{(1-f)+f n_{\mathrm{TM}}^2}+\sqrt{\frac{n_{\mathrm{TE}}^2}{n_{\mathrm{TE}}^2(1-f)+f}} \bigg)/2
\end{eqnarray}
for both TE and TM polarizations. Note that TE and TM polarizations mentioned in this paper are an approximation since the fields are not purely polarized in 3D structures. A more exact way to describe them is TE-like/TM-like, where electrical field is mostly parallel/vertical to the grooves \cite{Joannopoulos2008}. However, this is still an approximation because the waveguide is asymmetric so there is no horizontal mirror symmetric plane. The second-order terms used in Eq.~\ref{eq:1DEffectiveIndex1}\&\ref{eq:1DEffectiveIndex2} better approximate the effective index in the case that the wavelength is not very large comparing with size of unit cell, e.g. $\lambda=6a$ used in this paper. Most current metamaterial device designs are using the zeroth-order approximation only \cite{Valentine2009}, even when the unit cell size is not far smaller than the operational wavelength. This is fine for those devices where high accuracy results are not important. However, for devices such as L\"uneburg lens, all waves are focusing to a single point so light manipulation is more challenging. Therefore, more precise effective index prediction is needed and second-order corrections are included.

The dispersion relation of the effective guiding medium, i.e. the relationship between $k_z$ and $\omega$, is governed by the guidance condition of an asymmetric dielectric waveguide for both TE and TM polarizations \cite{Kong2008}


\begin{figure}[htbp]
\centering
\subfloat[]{ \includegraphics[width=6cm]{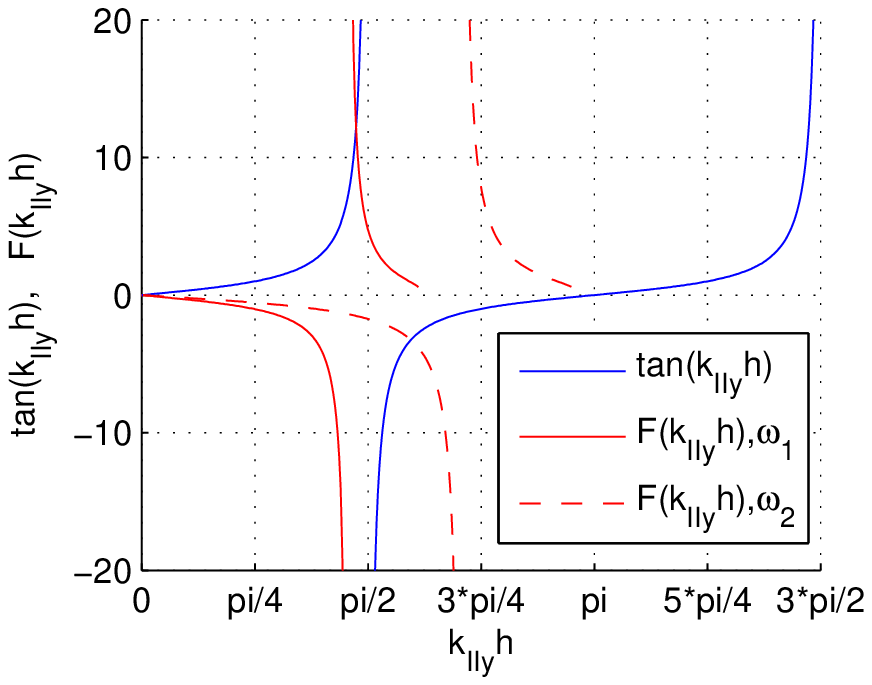} }
\subfloat[]{ \includegraphics[width=6cm]{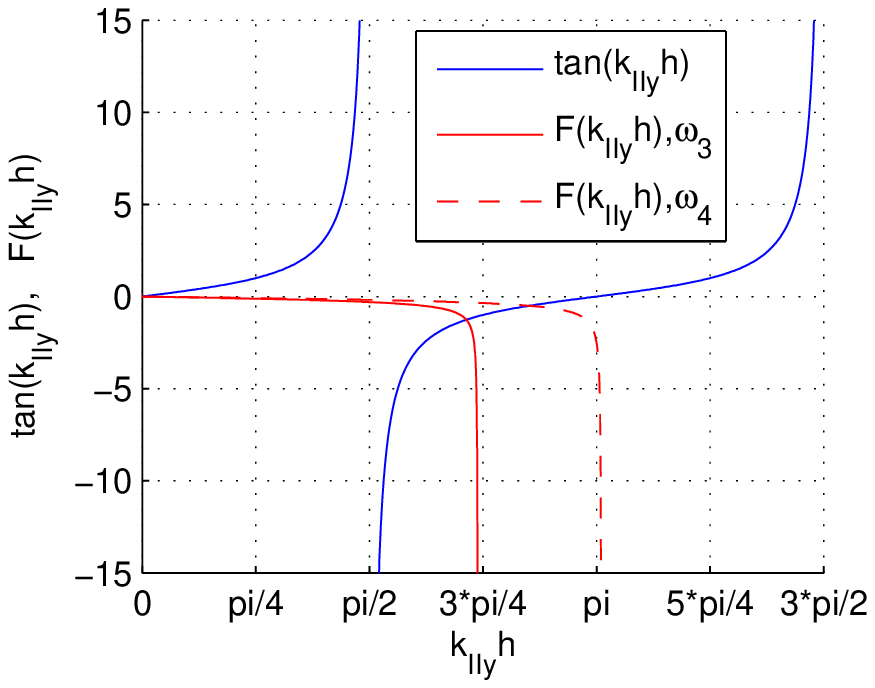} }
\caption{Graphical solutions of wave guidance condition (Eq.~\ref{eq:TEGuide}--\ref{eq:TMGuide}) for TE (a) and TM (b) polarizations. Blue and red lines are the left and right hand sides of these equations, respectively. Operating frequencies $\omega_1=0.11\times 2\pi c/a$, $\omega_2=0.16\times 2\pi c/a$, $\omega_3=0.14\times 2\pi c/a$ and $\omega_4=0.18\times 2\pi c/a$. Rod radius $r=0.50a$.}
\label{fig:GraphicSolution}
\end{figure}

\begin{eqnarray}
\mathrm{(TE:)~} \tan(k_{\mathrm{I\!I}y}h) &=& \frac{\epsilon_{\mathrm{I\!I}}{k_{\mathrm{I\!I}y} ( \epsilon_{\mathrm{I\!I\!I}}\sqrt{k_{z}^2-\epsilon_{\mathrm{I}}\omega^2/c^2} + \epsilon_{\mathrm{I}}\sqrt{k_{z}^2-\epsilon_{\mathrm{I\!I\!I}}\omega^2/c^2}} )} {\epsilon_{\mathrm{I}}\epsilon_{\mathrm{I\!I\!I}}k_{\mathrm{I\!I}y}^2 - \epsilon_{\mathrm{I\!I}}^2\sqrt{k_{z}^2 - \epsilon_{\mathrm{I}}\omega^2/c^2} \sqrt{k_{z}^2-\epsilon_{\mathrm{I\!I\!I}}\omega^2/c^2}} \equiv F_{\mathrm{TE}}(k_{\mathrm{I\!I}y}h),
\label{eq:TEGuide}
\end{eqnarray}
\begin{eqnarray}
\mathrm{(TM:)~} \tan(k_{\mathrm{I\!I}y}h) &=& \frac{{k_{\mathrm{I\!I}y} ( \sqrt{k_{z}^2-\epsilon_{\mathrm{I}}\omega^2/c^2} + \sqrt{k_{z}^2-\epsilon_{\mathrm{I\!I\!I}}\omega^2/c^2}} )} {k_{\mathrm{I\!I}y}^2 - \sqrt{k_{z}^2-\epsilon_{\mathrm{I}}\omega^2/c^2} \sqrt{k_{z}^2-\epsilon_{\mathrm{I\!I\!I}}\omega^2/c^2}} \equiv F_{\mathrm{TM}}(k_{\mathrm{I\!I}y}h),
\label{eq:TMGuide}
\end{eqnarray}
\noindent where $k_{z} = \sqrt{\epsilon_{\mathrm{I\!I}}\omega^2/c^2-k_{\mathrm{I\!I}y}^2}$ is the phase--matched propagation constant. These equations can be solved by a graphical method and an example is illustrated in Fig.~\ref{fig:GraphicSolution}.  It is observed that one and only one intersection is obtained for each frequency, meaning that only one fundamental mode is supported. Full dispersion relations $k_z(\omega)$ are shown in the following section.

\begin{figure}[htbp]
\centering
\subfloat[]{ \includegraphics[width=2cm]{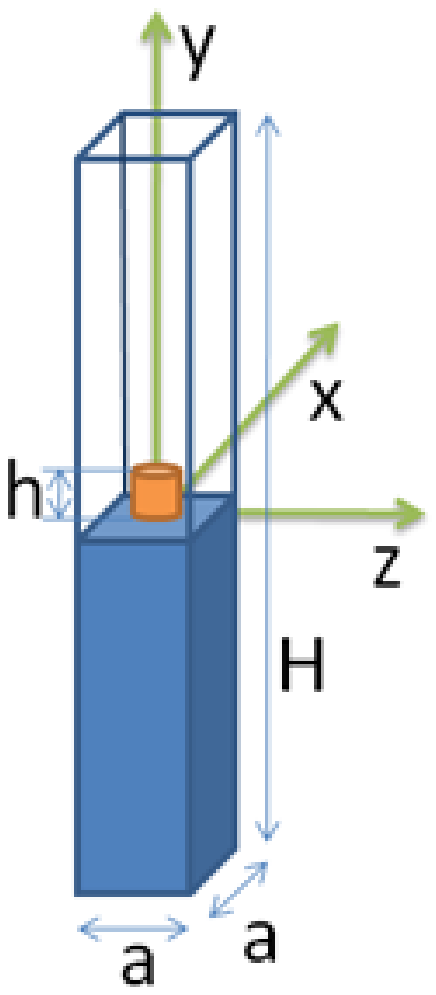} }
\subfloat[]{ \includegraphics[width=4.8cm]{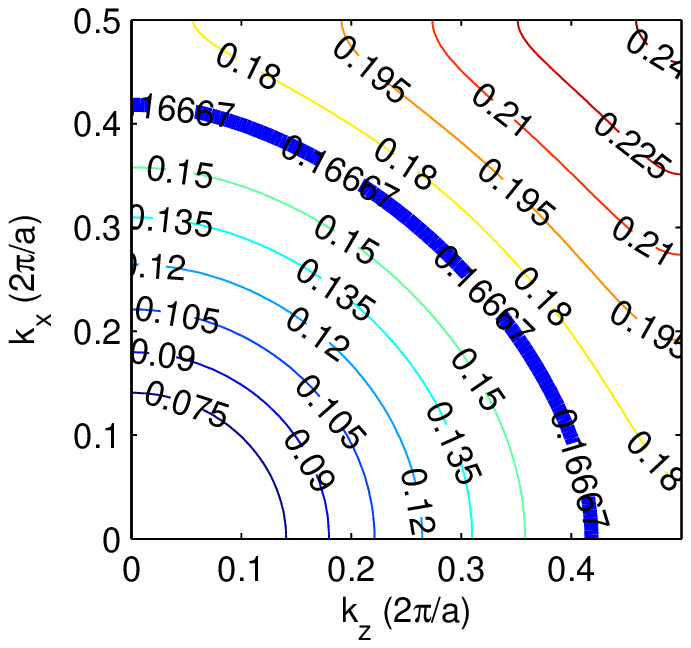} }
\subfloat[]{ \includegraphics[width=4cm]{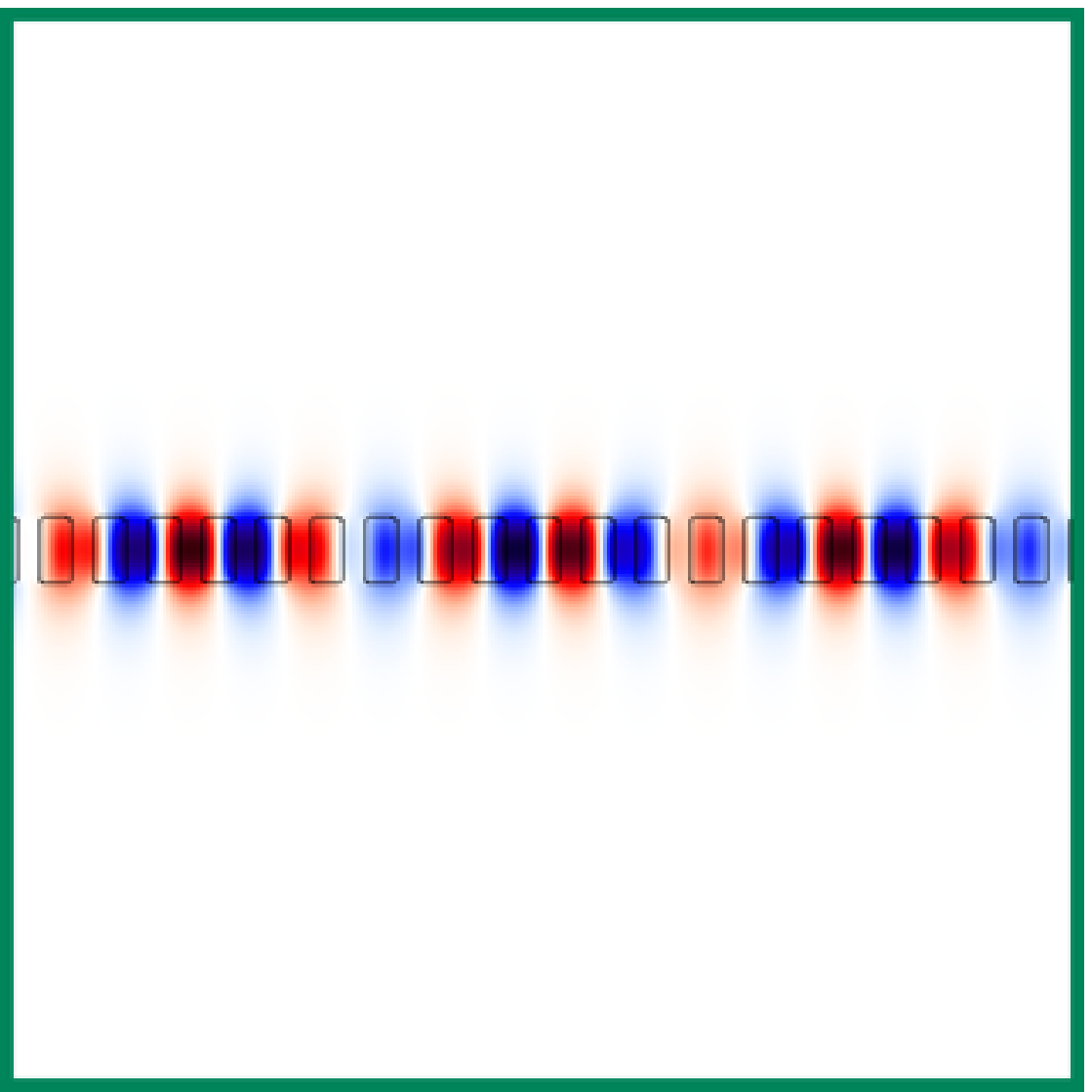} }
\caption{(a) The supercell used in the DBD method for the finite height rod lattice structure. (b) Isofrequency contour of the supercell with $r=0.50a$ where the first band only is shown. Labels on the lines denote the corresponding normalized frequency $\omega a/2\pi c$. The bold blue line corresponds to the wavelength $\lambda=6a$ used in this paper. (c) Field distribution of the waveguide slab at a particular $x$ slice. Color shading denotes magnetic field ($H_y$) distribution and black contours illustrate silicon rods.}
\label{fig:FirstMethod}
\end{figure}

The EGM method described above is compared with the conventional DBD method. To apply the DBD method, we need to calculate the band diagram of the 3D super cell shown in Fig.~\ref{fig:FirstMethod}(a). The supercell height is taken as large as $H=20a$ to better emulate the real structure of Fig.~\ref{fig:struct}(a), where the air and glass spaces tend to infinity. In other words, we seek to minimize the interference between neighboring unit cells along the vertical ($y$) direction. We used the MIT Photonic--Bands (MPB) mode solver \cite{Johnson2001} to calculate the dispersion diagram. In Fig.~\ref{fig:FirstMethod}(b--c) we show an example MPB result for our chosen lattice and the specific value $r=0.5a$, for temporal frequency $\omega = 1/6\times 2\pi c/a$. From Fig.~\ref{fig:FirstMethod}(b) we observe that for the chosen values of $r$ and $\omega$, the isofrequency contour \cite{Joannopoulos2008} is almost a circle, indicating that this unit cell is isotropic. Therefore, when using DBD in this particular geometry, it is sufficient to consider $k_z(\omega)$ only. However, this is not generally true in other geometries as $r$ or $\omega$ increase.

Figure~\ref{fig:FirstMethod}(c) shows the mode shape for the same geometry. It can be seen that the field is effectively concentrated near the silicon rod portion of the cell. The relative intensities at two horizontal cell boundaries $y=\pm H/2$ were $5.6\times 10^{-6}$ and $3.8\times 10^{-6}$ at the top and bottom, respectively, compared to the peak value that occurred at $y=159~\mathrm{nm}$ from the rod base. This validates our choice of $H$ as sufficiently large.


Comparing with the DBD method, the EGM method can provide deeper physical insights with all--analytical solutions, and is generally faster since it avoids solving numerical electromagnetic solutions in 3D.

\subsection{Effective refractive index and rod radius relationship}
\label{sec:rVSn}

In this section, the relationship between the effective refractive index and rod radius is calculated. The results of EGM method are compared with the ones obtained from DBD method.

\begin{figure}[htbp]
\centering
\subfloat[]{ \includegraphics[width=6.2cm]{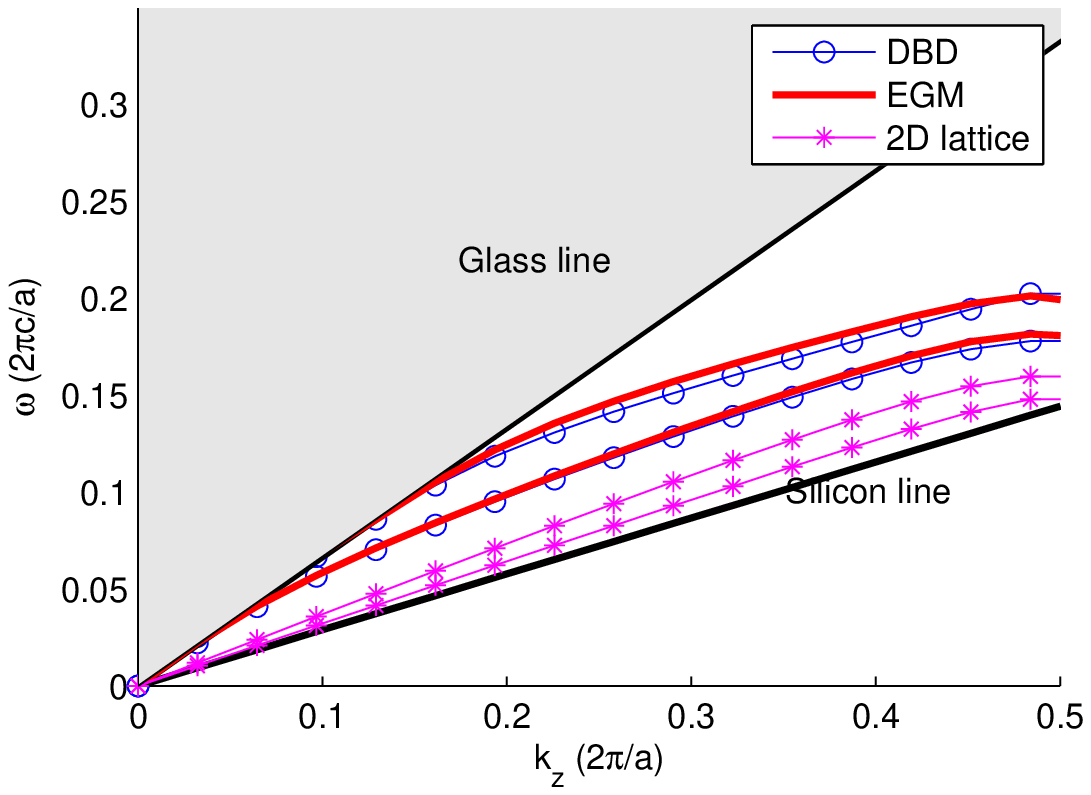} }
\subfloat[]{ \includegraphics[width=6.2cm]{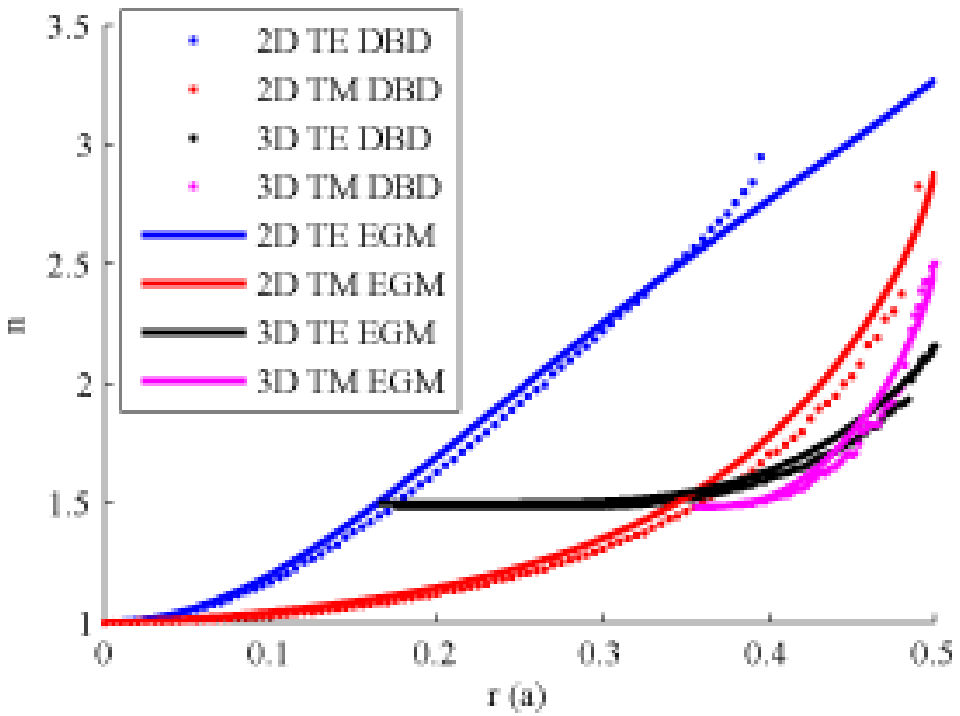} }
\caption{(a) Comparison between the dispersion relation for finite--height silicon rod lattice [Fig.~\ref{fig:struct}(a)] calculated from the EGM and DBD method, and the dispersion relation for infinite--height 2D rod lattice [Fig.~\ref{fig:struct}(b)]. For each case, the two lowest bands representing the TM and TE modes are shown. (b) Relationship between effective refractive index and rod radius calculated from both methods, compared with the relationship for infinite--height 2D rod lattice. Free space wavelength of light is $\lambda=6a=1550~\textrm{nm}$.
}
\label{fig:RadiusIndexRelation}
\end{figure}

Figure~\ref{fig:RadiusIndexRelation}(a) shows the dispersion relation of the finite--height rod lattice calculated with both DBD and EGM methods, as well as with the 2D (infinite rod height) assumption, for rod radius $r=0.5a$. Based on the dispersion relation, effective refractive indices for unit cells with different rod radii are calculated as $n_{\mathrm{eff}}=ck_z/\omega$, shown in Fig.~\ref{fig:RadiusIndexRelation}(b). The results given by the DBD and EGM methods are in good agreement with each other, with maximum percentage errors of 7.3\% and 6.0\% for 2D and 3D cases, respectively. It is observed that the effective refractive indices of the finite--height rods are significantly different than those assuming infinite height. This is to be expected due to weak guidance: as can been seen in Fig.~\ref{fig:FirstMethod}(c), a large portion of the field extends outside the rods to spaces of air and substrate. When the rod radii are below certain values ($0.17a$ for TE and $0.35a$ for TM), the propagation modes are not guided so the effective indices are not shown. The discontinuities observed in the 2D effective index curves for DBD method beyond certain values of rod radii ($0.40a$ for TE and $0.49a$ for TM) result from the emergence of a photonic crystal bandgap at these values. At this frequency range, even though the 2D infinite--height lattice is within the bandgap, the confined (slab waveguide) geometry is still propagating; this is because the light is mostly outside the dielectric region, so propagation takes place in the free space (hence the lower index). To calculate the propagation constant in this regime, we still need an effective index value and EGM provides it (it turns out to be large than 3, typically).


\section{Optimal design of the subwavelength L\"uneburg lens}

We re--design and numerically verify the subwavelength L\"uneburg lens \cite{Luneburg1944,Takahashi2010,Takahashi2011,Gao2011}, which was previously designed under 2D assumption. Here, we still design the L\"uneburg lens as a structure consisting of finite--height rods with adiabatically changing radius $r$ across the lattice of fixed constant $a$. At each coordinate $\rho$, we emulate the L\"uneburg distribution $n(\rho)=n_0\sqrt{2-(\rho/R)^2}$ by choosing the rod radius $r$ at coordinate $\rho$ from Fig.~\ref{fig:RadiusIndexRelation}(b) such that $n_{\mathrm{eff}}^{\mathrm{3D}} = n(\rho)$, as opposed to using $n_{\mathrm{eff}}^{\mathrm{2D}} = n(\rho)$. The design has to be carried out separately for the TE and TM polarizations. The ambient index is chosen as $n_0=1.53$.

\begin{figure}[htbp]
\centering
\subfloat[]{ \includegraphics[width=3cm]{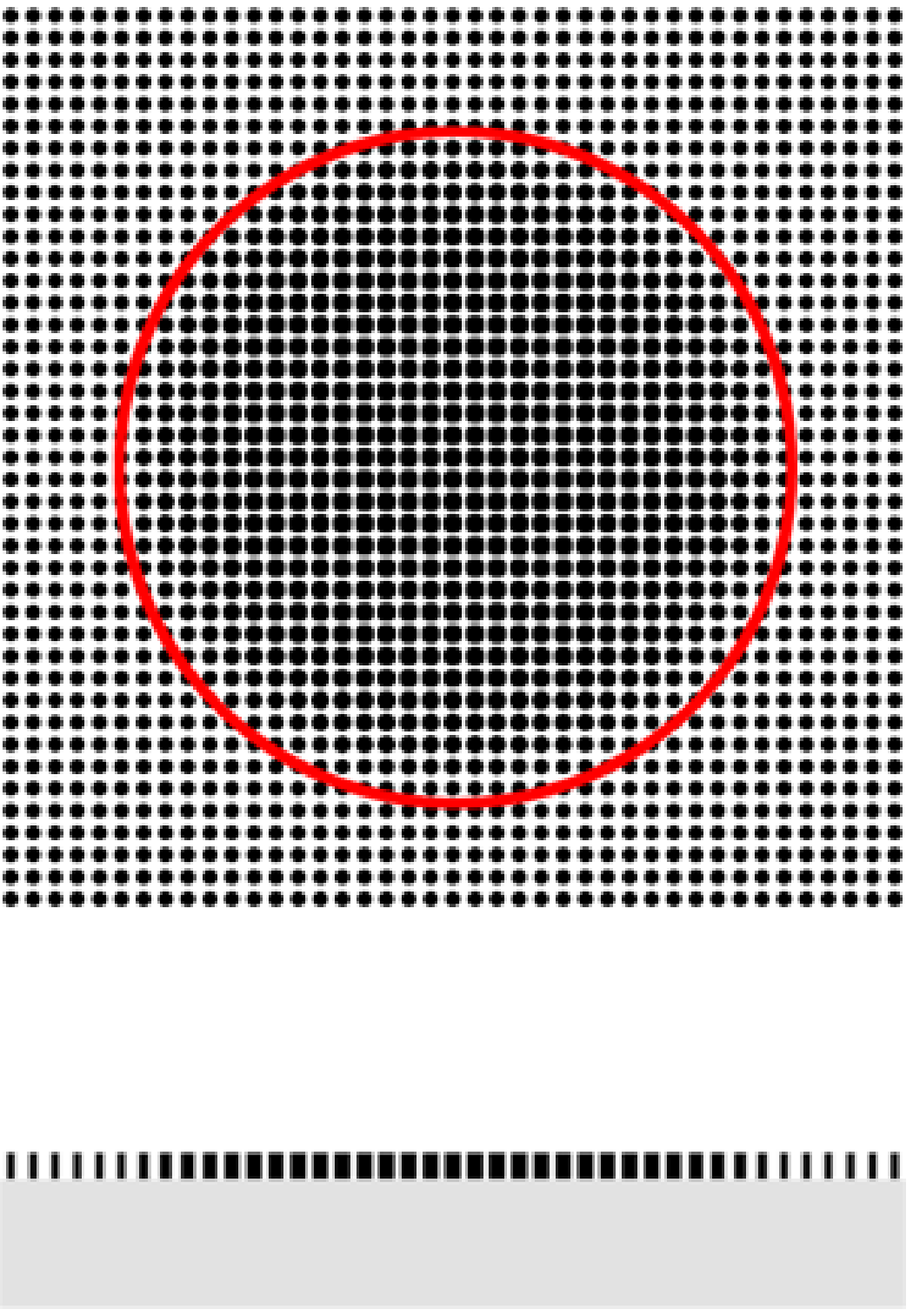}}
\subfloat[]{ \includegraphics[width=5.4cm]{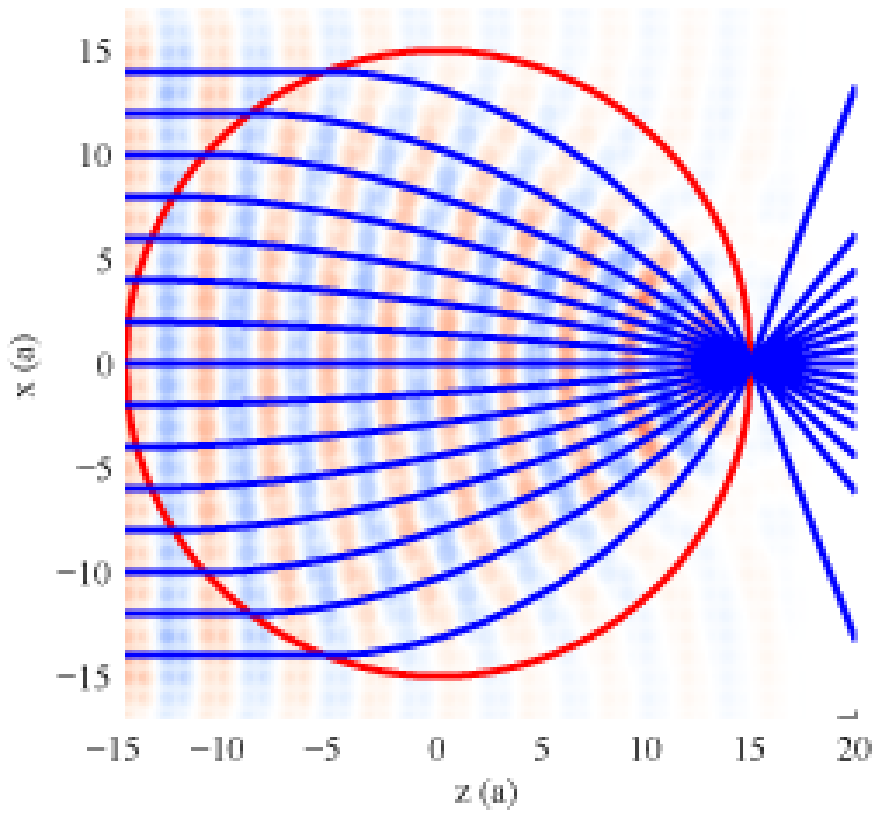} }

\subfloat[]{ \includegraphics[width=3cm]{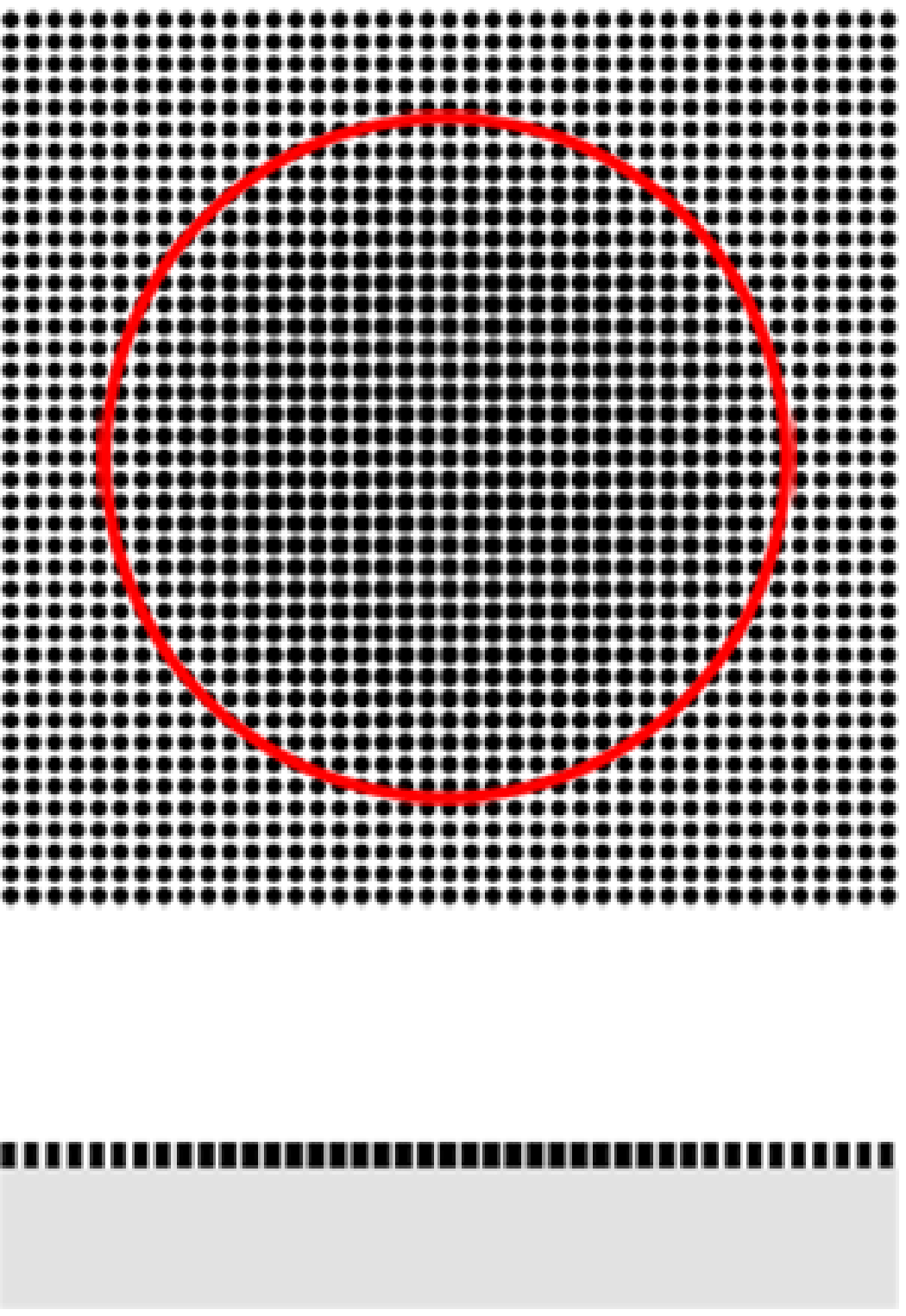}}
\subfloat[]{ \includegraphics[width=5.4cm]{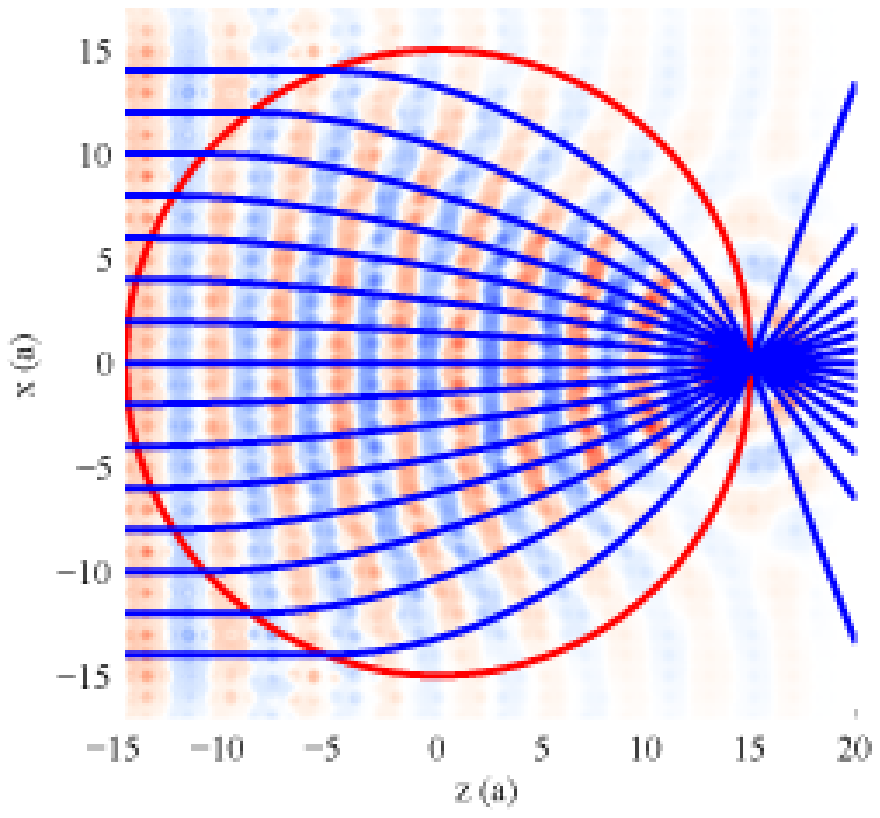} }
\caption{(a) Top view and side view of the thin--film subwavelength L\"uneburg lens designed by EGM method for TE mode and (b) the corresponding 3D FDTD and Hamiltonian ray tracing results. (c) Top view and side view for TM mode and (d) the corresponding 3D FDTD and ray tracing results. Red circles outline the edge of L\"uneburg lens, where radius $R=30a$. Blue lines are the ray tracing results and color shading denotes the field [$H_y$ for (b) and $E_y$ for (d)] distribution, where red is positive and blue is negative.}
\label{fig:WaveguideStructFDTD}
\end{figure}

\begin{figure}[htbp]
\centering
\subfloat[]{ \includegraphics[width=3cm]{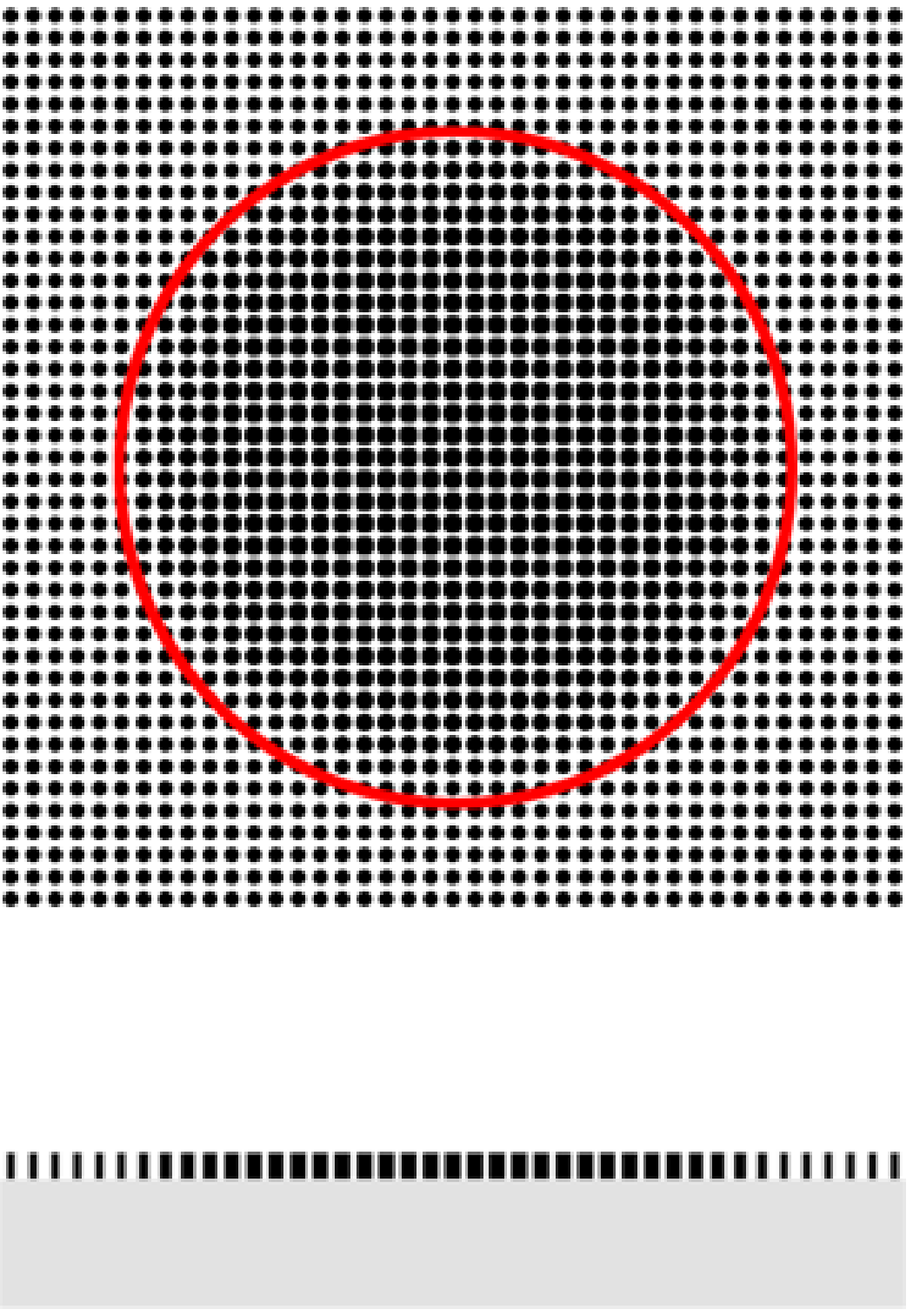}}
\subfloat[]{ \includegraphics[width=5.4cm]{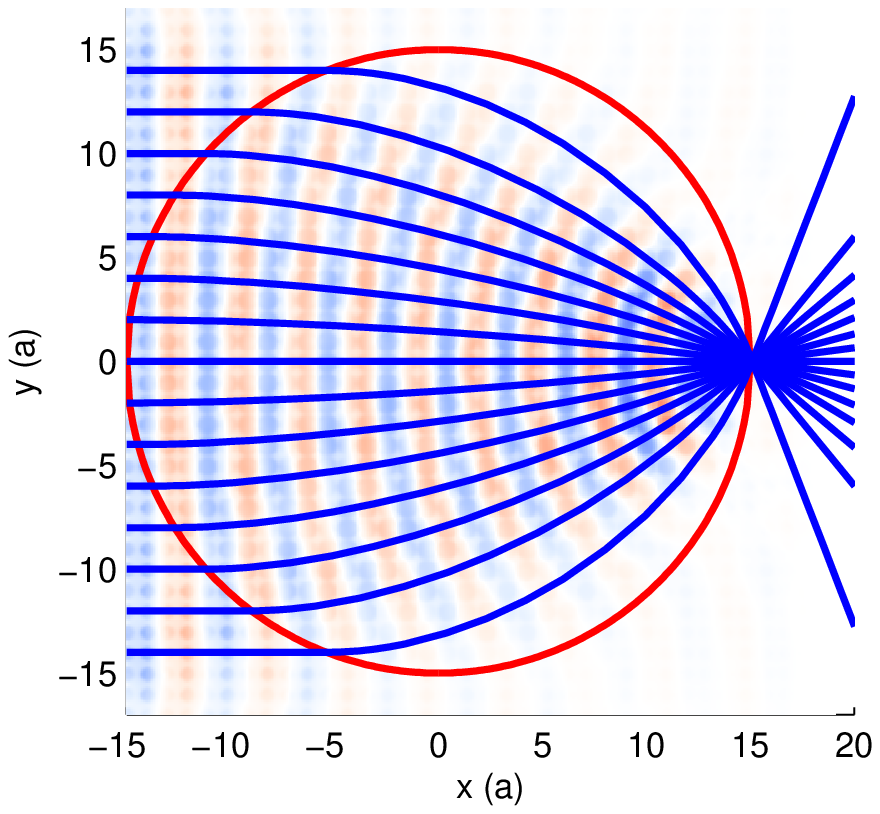} }

\subfloat[]{ \includegraphics[width=3cm]{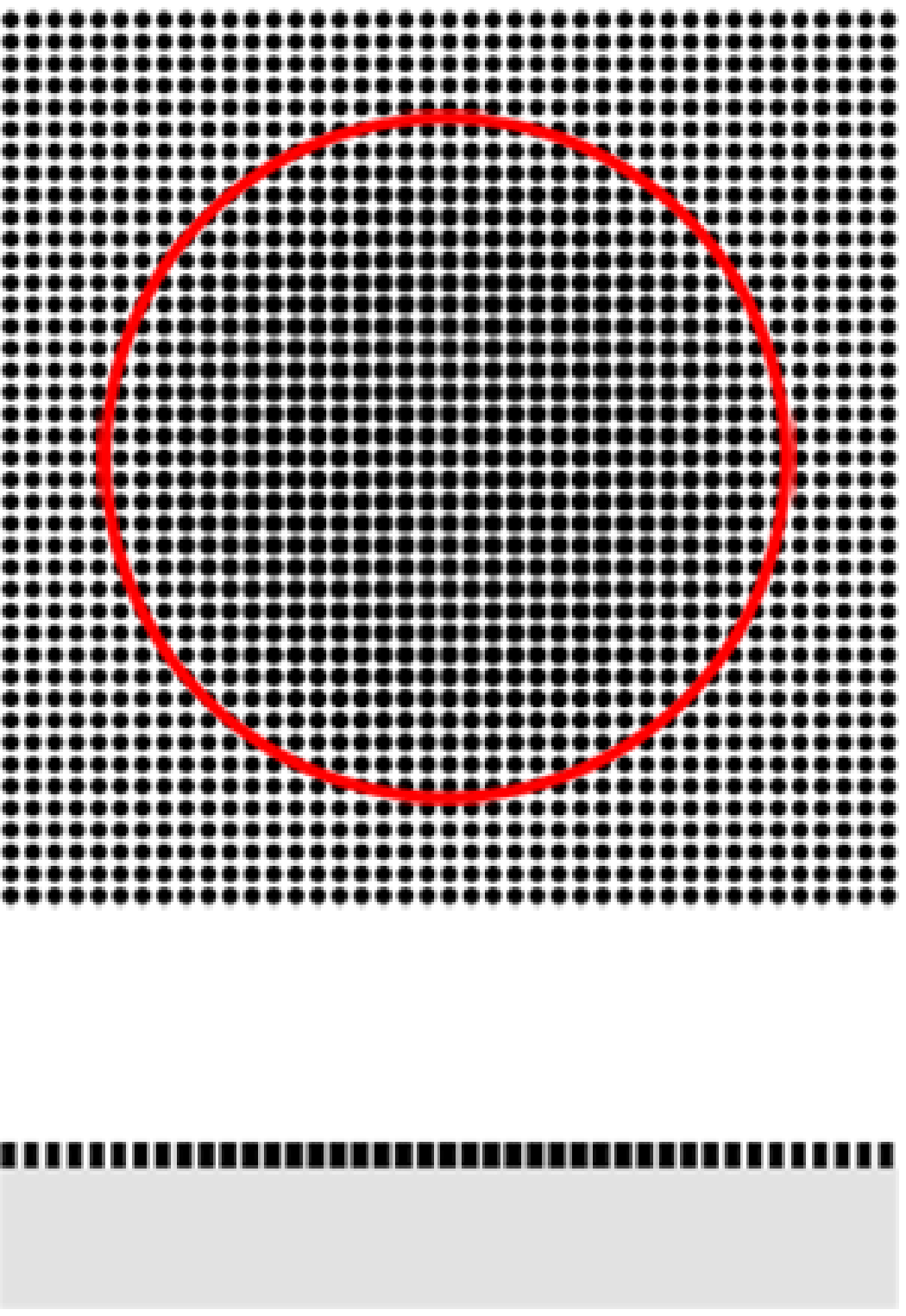}}
\subfloat[]{ \includegraphics[width=5.4cm]{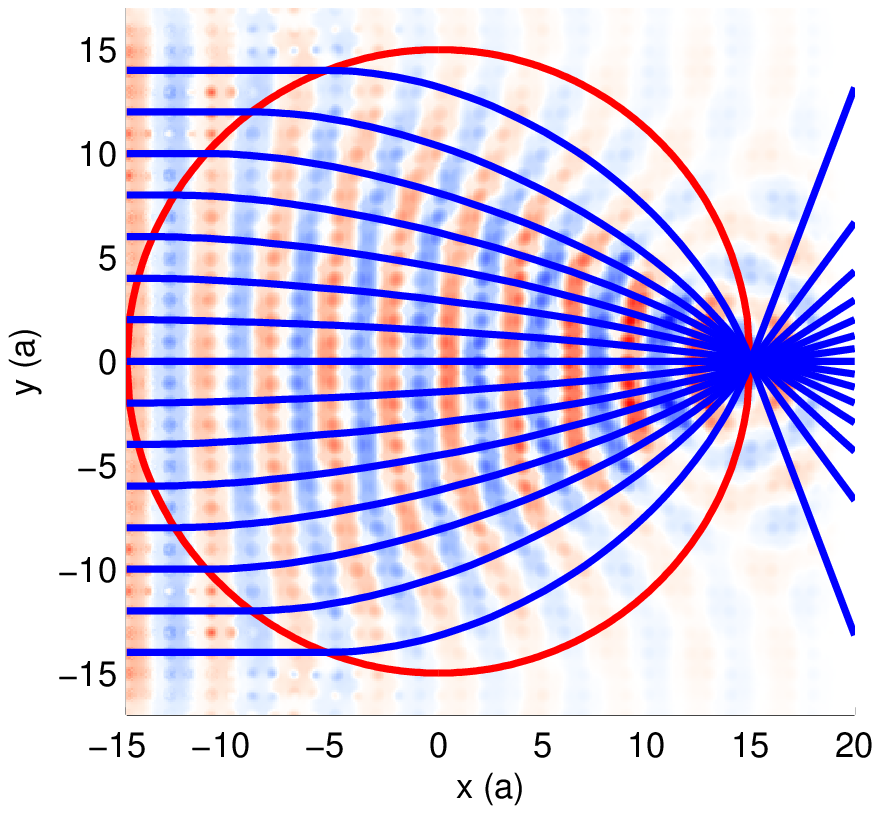} }
\caption{Structure and the corresponding 3D FDTD and Hamiltonian ray tracing for the thin--film subwavelength L\"uneburg lens shown in Fig.~\ref{fig:WaveguideStructFDTD}, but designed by the DBD method instead.}
\label{fig:WaveguideStructFDTDDBD}
\end{figure}

Figure~\ref{fig:WaveguideStructFDTD} illustrates the lens structures and the corresponding 3D finite--difference time--domain (FDTD) simulation results for the actual adiabatically variant thin--film nanostructured L\"uneburg lens performed by MIT Electromagnetic Equation Propagation (MEEP) \cite{Oskooi2010}. With plane wave illumination, almost diffraction--limited focal points at the edge can be observed for both TE and TM polarizations. For a more computationally efficient and intuitive representation we also ray--traced the field inside the L\"uneburg structure using the adiabatic Hamiltonian method \cite{Jiao2004,Russel1999}. The ray position $\textbf{q}$ and momentum $\textbf{p}$ are obtained by solving the two sets of coupled ordinary differential equations
\begin{eqnarray}
\frac{d\textbf{q}}{d\sigma}=\frac{\partial H}{\partial \textbf{p}},~~ \frac{d\textbf{p}}{d\sigma}=-\frac{\partial H}{\partial \textbf{q}},
\end{eqnarray}
\noindent where $H(\textbf{q},\textbf{p}) \equiv \omega (\rho, \textbf{k})$ is obtained from the dispersion diagram at each coordinate $|\textbf{q}|=\rho$ and for $\textbf{k} \equiv \textbf{p}$. Ray tracing results are superimposed in Fig.~\ref{fig:WaveguideStructFDTD} with FDTD results, and are seen to be in good agreement. Furthermore, as a comparison, similar thin--film L\"uneburg lens is designed using the DBD method and simulation results are shown in Fig.~\ref{fig:WaveguideStructFDTDDBD}. It is observed that results of the all--analytical EGM method design agree with those from the DBD method.

\begin{figure}[htbp]
\centering
\includegraphics[width=8cm]{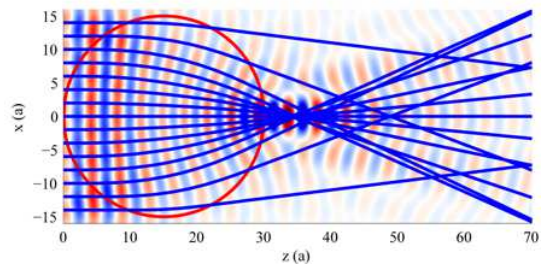}
\caption{FDTD and Hamiltonian ray--tracing results of the subwavelength L\"uneburg lens made of finite height silicon rods, but designed assuming infinite height. The color conventions are the same as in Fig.~\ref{fig:WaveguideStructFDTD}(b\&d)}
\label{fig:OriginalFDTD}
\end{figure}

To compare the redesigned lens (3D, finite height) with the original design (2D, infinite height), we repeated the design using the values of refractive indices predicted by the dispersion relation of the infinite--height rod lattice (see Fig.~\ref{fig:RadiusIndexRelation}(b) blue and red solid curves). In this case, we are forced to use TM polarization only because the TE polarization reaches the bandgap for relatively small value of $r$, not leaving enough room to implement the L\"uneburg profile with rod radius $r$ large enough to be robust to practical lithography and etching methods (in our experiment, this requires $r\ge 0.27a$ \cite{Takahashi2010,Takahashi2011}). It can be observed from the FDTD and Hamiltonian ray--tracing results shown in Fig.~\ref{fig:OriginalFDTD} that the focal point is outside the lens edge and it is strongly aberrated. This is in good agreement with the experimental results of the original design \cite{Takahashi2010,Takahashi2011}.

\section*{Acknowledgements}
The authors thank Lei Tian for useful discussions and Justin W. Lee for setting up the computation server. Financial support was provided by Singapore's National Research Foundation through the Singapore--MIT Alliance for Research and Technology (SMART) Centre and the Air Force Office of Scientific Research MURI program on Nanomembranes under contract No. FA9550-08-1-0379.

\end{document}